\documentstyle[12pt,epsfig,axodraw]{article}

\begin{document} 
\pagestyle{empty}

\begin{center} 
{\Large Neutrinos masses in a supersymmtric model with exotic right-handed neutrinos in Global $(B-L)$ Symmetry}
\end{center}

\begin{center}
M. C. Rodriguez \\
{\it Grupo de F\'\i sica Te\'orica e Matem\'atica F\'\i sica \\
Departamento de F\'\i sica \\
Universidade Federal Rural do Rio de Janeiro - UFRRJ \\
BR 465 Km 7, 23890-000 \\
Serop\'edica, RJ, Brazil, \\
email: marcoscrodriguez@ufrrj.br \\} 
\end{center} 

\begin{abstract}
We build a supersymmetric version with $SU(3)_{C}\otimes SU(2)_{L}\otimes U(1)_{Y}$ gauge symmetry. We impose, also, the ${\cal Z}_{3}$ symmetry to keep the proton stable at least at tree level. The model has three right-handed neutrinos with non identical $(B-L)$ charges and some extra Higgs 
fields. There are, also, a global $(B - L)$, where ($B$) and ($L$) are the usual baryonic and leptonic numbers respectivelly, symmetry. We will show only right handed neutrinos and one left-handed neutrino get mass at tree level while the others two left handed neutrinos get their masses at 1-loop level. We will also explain the mixing angle in the neutrino sector in agreement with the experimental data.

\end{abstract}

Keywords: Extensions of the electroweak gauge sector; 
Supersymmetric models; Neutrino mass and mixing; 
Non-standard-model neutrinos, right-handed neutrinos, 
etc.\\ 
PACS number(s): 12.60.Cn; 12.60. Jv; 14.60.Pq; 14.60.St.

\section{Introduction}
\label{sec:intro}

Today we know that neutrinos have mass and we notice oscillation at neutrinos sector, for example in the case of solar neutrinos problem is explained as $\nu_{e}$ disappearance while  to explain atmospheric neutrino data via $\nu_{\mu}\rightarrow \nu_{\tau}$ oscillation and the experimental data has two large angles, they are $\theta_{solar}$ and $\theta_{atm}$ 
and such mixing is termed as ``tribimaximal" mixing scheme.
 
There is an interesting model with the same gauge symmetry as in the Standard Model (SM), it means
$SU(3)_{C}\otimes SU(2)_{L}\otimes U(1)_{Y}$ gauge symmetries and we introduce two right-handed 
neutrinos having $(B-L)=-4$ and the third one having 
$(B-L)=5$ and besides the usual scalar $S$ we introduce two extra dublets (they are inert under ${\cal Z}_{3}$ symmetry), this model 
was presented at \cite{Machado:2017ksd}, in this 
reference the authors also presented a first short analyse about the 
supersymmetric version of this model. Our goal here is present this model in more detail.

We will present the supersymmetric version of the model described at~\cite{Machado:2017ksd}, where the particle spectra is 
enlarged with four inert doublets, some scalars as singlets under $SU(2)$ symmetry and three right-handed neutrinos with 
non-standard assignment of global $(B-L)$ symmetry which make the model 
anomaly free~\cite{Montero:2007cd}, the SUSY versions of these models 
were done at \cite{Montero:2016qpx,Rodriguez:2020fvo}, and we will 
also impose ${\cal Z}_{3}$ symmetry to avoid the 
proton decay at 
least at tree level.

The outline of this paper is as follows: In Sec.(\ref{sec:model}) we 
present the particle content of this model in superfield formalism, in 
Sec.(\ref{sec:lagrangianm1}) we present the lagrangian of this model. In 
Sec.(\ref{sec:fermionmasses}) we calculate the masses to fermions at tree level. We present, in a short way, the scalar potential at Sec.(\ref{sec:scalarpotential}). Some left-handed neutrinos get mass their masses at to 1-loop level 
in Sec.(\ref{sec:1loopmechanism}) in similar way as showed at 
\cite{Machado:2017ksd}. In the end of this article, we present our conclusion.

\section{The Model}
\label{sec:model}

Now we will review the non-SUSY version of the model, 
where our symmetry is 
\begin{eqnarray}
SU(3)_{C}\otimes SU(2)_{L}\otimes U(1)_{Y},
\end{eqnarray}
as the gauge symmetry of this model is the same as in the Standard Model (SM), we have the 
same gauge bosons of SM. They are the gluons fields $g^{a}$, and the bosons $W^{i}$ of the group $SU(2)$ and $b^{\prime}$ of $U(1)$ and we have ommited the Lorentz indeces. We will, also, impose $(B-L)$ as global symmetry.

The representation content of the non-SUSY model is 
the following: under $SU(2)_{L}$ we have the lepton doublets 
$L_{iL}=(\nu_{i}, l_{i})^{T}_{L}\sim ({\bf 1},{\bf 2},-1)$\footnote{The parentheses are the transformation properties under the respective representation of $(SU(3)_{C},SU(2)_{L},
U(1)_{Y})$.}, and $(i=1,2,3)$ denote 
fermion generations with $(B-L)=-1$; charged singlets 
$E_{iR}\sim ({\bf 1},{\bf 1},-2)$ with $(B-L)=-1$; three sterile neutrinos, one 
of them $N_{1R}\sim ({\bf 1},{\bf 1},0)$ with $(B-L)=-5$ and the others two 
$N_{\alpha R}\sim ({\bf 1},{\bf 1},0),\, ( \alpha =2,3)$ with $(B-L)=+4$; the quarks 
sector is exactly the same as in the SM with 
$(B-L)=(+1/3)$; we also 
introduce the SM Higgs 
$S \sim ({\bf 1},{\bf 2},1)$ 
with $(B-L)=0$, and two scalar 
inert doublet $D_{1,2}\sim ({\bf 1},{\bf 2},1)$ and $(B-L)=+6,-3$, respectively, 
more details about this model can be find at \cite{Machado:2017ksd}. 

Now we will start to construct the supersymmetric version of this model, all the fields listed above we will put, as usual in supersymmetric 
models, in chiral superfield, the gauge sector of this model is identical of SM are introduced at vector superfield
~\cite{Rodriguez:2019,Rodriguez:2020,dress,Baer:2006rs,
ait1,Rodriguez:2016esw}.

\subsection{The Superfields}

The usual fermions as done in Minimal Supersymmetric Standard Model (MSSM) are $\hat{L}_{iL},\hat{E}_{iR},\hat{Q}_{iL},\hat{u}_{iR}$ and $\hat{d}_{iR}$ and the news one are $\hat{N}_{1R}$ and $\hat{N}_{\beta R}$ are put in chiral superfields 
and their quantum numbers are shown at 
Tabs.(\ref{usuallepton},\ref{usualquark}).
\begin{table}[h]
\begin{center}
\begin{tabular}{|c|c|c|c|c|}
\hline 
${\rm{Superfield}} $ & $\hat{L}_{iL}$ & $\hat{E}_{iR}$ & $\hat{N}_{1R}$ & $\hat{N}_{\beta R}$ \\
\hline 
$SU(3)_{C}\otimes SU(2)_{L}\otimes U(1)_{Y}$ & $({\bf 1}, {\bf 2},-1)$ & 
$({\bf 1}, {\bf 1},2)$ & $({\bf 1}, {\bf 1},0)$ & $({\bf 1}, {\bf 1},0)$  \\
\hline 
$(B-L)$ & $(-1)$ & $(+1)$ & $(+5)$ & 
$(-4)$   \\
\hline
\end{tabular}
\end{center}
\caption{Transformation properties of the lepton under $(SU(3)_{C},SU(2)_{L},
U(1)_{Y})$ and $(B-L)$.}
\label{usuallepton}
\end{table}

\begin{table}[h]
\begin{center}
\begin{tabular}{|c|c|c|c|}
\hline 
${\rm{Superfield}} $ & $\hat{Q}_{iL}$ & $\hat{u}_{iR}$ & $\hat{d}_{iR}$ \\
\hline 
$SU(3)_{C}\otimes SU(2)_{L}\otimes U(1)_{Y}$ &  
$\left({\bf 3}, {\bf 2}, \frac{1}{3} \right)$ & 
$\left({\bf 3^{*}}, {\bf 1},- \frac{4}{3} \right)$ & 
$\left({\bf 3^{*}}, {\bf 1}, \frac{2}{3} \right)$   \\
\hline 
$(B-L)$ &  $\left( + \frac{1}{3} \right)$ & 
$\left( - \frac{1}{3} \right)$ & $\left( - \frac{1}{3} \right)$  \\
\hline
\end{tabular}
\end{center}
\caption{Transformation properties of the quark under $(SU(3)_{C},SU(2)_{L},
U(1)_{Y})$ and $(B-L)$.}
\label{usualquark}
\end{table}

We add also three right-chiral neutrinos superfields, in similar way as 
done at MSSM with three right-handed neutrinos 
(MSSM3RHN) \cite{Rodriguez:2020}, we will represent them as  $\hat{N}_{iR}$
\footnote{It means $\hat{N}_{1R},\hat{N}_{2R}$ and $\hat{N}_{3R}$ and we will use this notation allways we do not says their $(B-L)$ charges.}, 
see~\cite{Rodriguez:2020,Rodriguez:2016esw}, in the following way 
\begin{eqnarray}
\hat{N}_{iR}(y, \theta )&=& \tilde{N}_{iR}(y)+ \sqrt{2} \left( \theta N_{iR}(y) \right) + 
\left( \theta \theta \right) F_{N_{iR}}(y), \,\ (i=1,2,3), 
\label{chiralrighthanedneutrinos}
\end{eqnarray}
where the fields $N_{iR}\equiv \nu^{c}_{iL}$, as we presented in 
some preliminar studies presented at 
~\cite{Rodriguez:2019,Rodriguez:2020}, are the right-handed 
neutrinos known as sterile neutrinos 
\cite{Volkas:2001zb} and we should also introduce three 
right-handed sneutrnios $\tilde{N}_{iR}$, and we will call them 
as sterile sneutrinos, as we have discussed 
at~\cite{Rodriguez:2020}. 

The new scalars, the left-handed sneutrinos 
$\tilde{\nu}_{iL}$ and the right-handed sneutrinos $\tilde{N}_{iR}$, 
are innert due $(B-L)$ symmetry, therefore, we can write the following constraints in their vaccum expectation values (vev)
\begin{equation}
\langle \tilde{\nu}_{iL} \rangle = 
\langle \tilde{N}_{1R} \rangle =
\langle \tilde{N}_{\beta R} \rangle =  0.
\label{vevextradoublets}
\end{equation}

An interesting feature of this model is that the right-handed 
neutrino $N_{1}$  do not mix with the right-handed neutrinos 
$N_{2R},N_{3R}$
\footnote{The same is hold to right-handed sneutrinos}
because they have differents $(B-L)$ quantum numbers, we will present it at Sec.(\ref{sec:neutrinomasses}). 

In the non-SUSY model, as we mentioned in our Introduction, the 
right-handed neutrinos $N_{1R},N_{\beta R}$ 
has $(B-L)$ charge as $-5,+4$, see \cite{Machado:2017ksd}, respectivelly. However in the 
supersymmetric model they are introduced as antiparticles, 
see Eq.(\ref{chiralrighthanedneutrinos}), due this 
fact they have opposite $(B-L)$ charge, as we presented at 
Tab.(\ref{usuallepton}), more details about it can be found at
~\cite{Rodriguez:2019,Rodriguez:2020}.

The scalars in dublets representation of this model are presented at 
Tabs.(\ref{scalardoublet},\ref{newscalardoublet})
\begin{table}[h]
\begin{center}
\begin{tabular}{|c|c|c|c|c|}
\hline 
${\rm{Superfield}} $ & $\hat{H}_{1}$ & $\hat{H}_{2}$ & $\hat{D}_{1}$ &  $\hat{D}_{2}$  \\
\hline 
$SU(3)_{C}\otimes SU(2)_{L}\otimes U(1)_{Y}$ & $({\bf 1}, {\bf 2},1)$ & $({\bf 1}, {\bf 2},-1)$ & 
$({\bf 1}, {\bf 2},1)$ & 
$({\bf 1}, {\bf 2},1)$  \\
\hline 
$(B-L)$ & $(0)$ & $(0)$ & $(-4)$ & 
$(+5)$  \\
\hline
\end{tabular}
\end{center}
\caption{Transformation properties of the scalars doublets introduced in the non-SUSY model under $(SU(3)_{C},SU(2)_{L},
U(1)_{Y})$ and $(B-L)$.}
\label{scalardoublet}
\end{table}

\begin{table}[h]
\begin{center}
\begin{tabular}{|c|c|c|}
\hline 
${\rm{Superfield}} $ & $\hat{D}^{\prime}_{1}$ & $\hat{D}^{\prime}_{2}$ \\
\hline 
$SU(3)_{C}\otimes SU(2)_{L}\otimes U(1)_{Y}$ & $({\bf 1}, {\bf 2},-1)$ & 
$({\bf 1}, {\bf 2},-1)$ \\
\hline 
$(B-L)$ & $(+4)$ & $(-5)$ \\
\hline
\end{tabular}
\end{center}
\caption{Transformation properties of the new 
scalars, introduced to avoid chiral anomalies in this model, under $(SU(3)_{C},SU(2)_{L},
U(1)_{Y})$ and $(B-L)$.}
\label{newscalardoublet}
\end{table}
The usual scalars $H_{1,2}$\footnote{In the non-SUSY model the scalar 
field was denoted as $S$ \cite{Machado:2017ksd}} their vev, as usual, 
are given by:
\begin{equation}
\langle H_{1} \rangle = \frac{v_{1}}{\sqrt{2}}, \,\ 
\langle H_{2} \rangle = \frac{v_{2}}{\sqrt{2}}.
\label{vevh1h2}
\end{equation}
while the new scalars $D_{1},D_{2}$ has $(B-L)$ charge as $+6,-3$, in 
the non-SUSY model, respectively. Those quantum number 
came from the following Yukawa interactions 
\begin{equation}
G_{i} \left( \overline{L^{c}}_{iL}D_{1} \right) N_{1R}+ 
P_{i \beta} \left( \overline{L^{c}}_{iL}D_{2} \right) N_{\beta R},
\end{equation}
where we have, as done usually in supersymmetric models defined
\begin{equation}
\left( AB \right) \equiv \epsilon_{\alpha \beta}A_{\alpha}B_{\beta},
\label{contracaodedubletos}
\end{equation} 
where $\alpha =1,2$ is a spinorial index and the Yukawa interactions are presented at 
\cite{Machado:2017ksd}. However our right-handed neutrinos have 
different $(B-L)$ charge and it imply our new scalars $D_{1},D_{2}$ have 
others $(B-L)$ charges. The motivation to those values will be more clear 
when we presente our superpotential at Eq.(\ref{mssmrpc}). 

The new scalars $D_{1},D^{\prime}_{1},D_{2},D^{\prime}_{2}$ are 
innert due ${\cal Z}_{3}$ symmetry, therefore, we can write
\begin{equation}
\langle D_{1} \rangle = \langle D_{2} \rangle = \langle D^{\prime}_{1} \rangle =
\langle D^{\prime}_{2} \rangle = 0.
\label{vevextradoublets}
\end{equation}
As a consequence, this fact imply that our gauge bosons have the 
following masses \cite{Rodriguez:2019,dress,Baer:2006rs,ait1,Rodriguez:2016esw}
\begin{eqnarray}
M_{W}= \left( \frac{gv_{2}}{2}\right) \sqrt{1+ \tan^{2}\beta}, 
\,\
M_{Z}= \frac{M_{W}}{\cos \theta_{W}},
\end{eqnarray}
where 
\begin{eqnarray}
\tan \beta = \left( \frac{v_{1}}{v_{2}}\right), \,\
\tan \theta_{W}= \left( \frac{g^{\prime}}{g}\right).
\label{betaparameter} 
\end{eqnarray}
The new parameter $\beta$ is a free parameter and $\theta_{W}$ is, as 
usual, the Weinberg angle.

In the non-SUSY model we can write the following Majorana mass term to our right-handed neutrinos \cite{Machado:2017ksd}
\begin{eqnarray}
M_{1}\overline{\left( N_{1R} \right)^{c}}N_{1R}+
M_{\alpha \beta}\overline{\left( N_{\alpha R} \right)^{c}}N_{\beta R}+hc,
\end{eqnarray} 
this term in terms of superfield would be translated as
\begin{eqnarray}
M_{1}\hat{\bar{N}}_{1R}\hat{N}_{1R}+
M_{\alpha \beta}\hat{\bar{N}}_{\alpha R}\hat{N}_{\beta R}+hc,
\end{eqnarray}
but their result is not a chiral superfield \cite{Rodriguez:2019,Rodriguez:2020} and as consequence 
this kind of term is not allowed in our superpotential, see Sec.(\ref{subsec:spotential}). 

Therefore, to obtain an arbitrary mass matrix for the neutrinos in this model we need to introduce 
some scalars in the singlet representation, as we present at 
Tab.(\ref{scalarsinglet}).
\begin{table}[h]
\begin{center}
\begin{tabular}{|c|c|c|}
\hline 
${\rm{Superfield}} $ & $\hat{\varphi}$ & $\hat{\phi}$   \\
\hline 
$SU(3)_{C}\otimes SU(2)_{L}\otimes U(1)_{Y}$ & $({\bf 1}, {\bf 1},0)$ & $({\bf 1}, {\bf 1},0)$  \\
\hline 
$(B-L)$ & $(-10)$ & $(+8)$  \\
\hline
\end{tabular}
\end{center}
\caption{Transformation properties of the scalars under $(SU(3)_{C},SU(2)_{L},
U(1)_{Y})$ and $(B-L)$.}
\label{scalarsinglet}
\end{table}
Their vev are
\begin{equation}
\langle \varphi \rangle = \frac{u_{1}}{\sqrt{2}}, \,\ 
\langle \phi \rangle = \frac{u_{2}}{\sqrt{2}}.
\label{vevextrasinglets}
\end{equation}
The extra scalar field $S$ is introduced to solve the $\mu$-problem, see~\cite{Rodriguez:2019,dress,Baer:2006rs,Rodriguez:2016esw}.

As happen with neutrinos, the higgsinos $\tilde{H}_{1,2}$ do not 
mix with others Higgsinos, but they mix with the gauginos in the 
same way as in the MSSM, 
because they have differents $(B-L)$ quantum numbers. The higgsinos 
$\tilde{D}_{1}$ and $\tilde{D}^{\prime}_{1}$ can mix, the same 
happen to $\tilde{D}_{2}$ and $\tilde{D}^{\prime}_{2}$. Therefore 
the higgsinos $\tilde{\varphi}, \tilde{\phi}$ and $\tilde{S}$ are 
already mass eigenstates and they are very massive, due their 
masses came from soft terms as we will present at 
Sec.(\ref{sec:lagsofttermsm1}).

Concerning the gauge bosons and their superpartners, known as gauginos, are introduced in vector superfields~
\cite{Rodriguez:2019,Rodriguez:2020,dress,Baer:2006rs,ait1,Rodriguez:2016esw}. See Tab.(\ref{table3}) where we we presente the particle content together 
with the gauge coupling constant of each group.

\begin{table}[h]
\begin{center}
\begin{tabular}{|c|c|c|c|c|c|}
\hline
${\rm{Group}}$ & ${\rm Superfield}$ & ${\rm{Bosons}}$ & ${\rm{Gaugino}}$ & ${\rm{Auxiliar \,\ Field}}$ & ${\rm constant}$ \\
\hline
$SU(3)_{C}$ & $\hat{G}^{a}$ & $g^{a}_{m}$ & $\tilde{g}^{a}$ & $D_{g}$ & $g_{s}$  \\
\hline
$SU(2)_{L}$ & $\hat{W}^{i}$ & $W^{i}_{m}$ & $\tilde{W}^{i}$ & $D_{W}$ & $g$ \\
\hline
$U(1)_{Y}$ & $\hat{b}^{\prime}$ & $b^{\prime}_{m}$ & $\tilde{b}^{\prime}$ & $D^{\prime}$ & $g^{\prime}$ \\
\hline
\end{tabular}
\end{center}
\caption{Information on fields contents of each vector superfield of this model. The Latin index $m$ identify Lorentz index 
as~\cite{Rodriguez:2019,Rodriguez:2020}}
\label{table3}%gaugefieldscontent
\end{table}

These are the minimal fields, we need to construct this supersymmetric model.

\subsection{R-Parity}
\label{Rparitym2}

Let us begining defining the $R$-parity in the model with the particle content listed above. We define at Tabs.(\ref{rcharge},\ref{rchargequarks}) the $R$-charge 
($n_{\Phi}$)\footnote{Here $\Phi$ means chiral superfield as defined at \cite{Rodriguez:2019,Rodriguez:2020,dress,Baer:2006rs,ait1,Rodriguez:2016esw}.} of each superfield in our model. Using these 
$R$-charges we can get the $R$-Parity of each fermion field contained in these chiral superfield, as shown at \cite{Rodriguez:2019,Rodriguez:2020}, these 
results we shown at Tab.(\ref{fermionblrm2}).
\begin{table}[h]
\begin{center}
\begin{tabular}{|c|c|c|c|c|c|}
\hline 
${\rm{Superfield}}$ & $\hat{L}_{iL}$ & $\hat{E}_{iR}$ & $\hat{N}_{1R}$ & $\hat{N}_{\beta R}$ & $\hat{S}$  \\
\hline 
$R-{\rm{charge}}$ & $n_{L}=(+1)$ & $n_{E}=(-1)$ & $n_{N_{1}}=(-1)$ & $n_{N_{\beta}}=(-1)$ & $n_{S}=(0)$    \\
\hline
\end{tabular}
\end{center}
\caption{Information about the $R$-charge ($n_{\Phi}$) of all the leptons and scalars at chiral superfields of this model, our notation here 
$S=H_{1,2},D_{1,2},D^{\prime}_{1,2},\varphi , \phi$.}
\label{rcharge}
\end{table}
\begin{table}[h]
\begin{center}
\begin{tabular}{|c|c|c|c|}
\hline 
${\rm{Superfield}}$ & $\hat{Q}_{iL}$ & $\hat{u}_{iR}$ & $\hat{d}_{iR}$  \\
\hline 
$R-{\rm{charge}}$ & $n_{L}=(+1)$ & $n_{u}=(-1)$ & $n_{d}=(-1)$ \\
\hline
\end{tabular}
\end{center}
\caption{Information about the $R$-charge ($n_{\Phi}$) of quarks at chiral superfields of this model.}
\label{rchargequarks}
\end{table}
\begin{table}[h]
\begin{center}
\begin{tabular}{|c|c|c|c|c|c|c|c|c|c|}
\hline 
${\rm{Fermion}} $ & $N_{1}$ & $N_{\beta}$ & $\tilde{D}_{1}$ & $\tilde{D}^{\prime}_{1}$ & $\tilde{D}_{2}$ & $\tilde{D}^{\prime}_{2}$ & 
$\tilde{\varphi}$ & $\tilde{\phi}$ & $\tilde{S}$     \\
\hline 
$(B-L)$ & $5$ & $-4$ & $-4$ & $+4$ & $+5$ & $-5$ & $-10$ & $+8$ 
& $0$     \\
\hline 
$R-{\rm{Parity}}$ & $+1$ & $+1$ & $-1$ & $-1$ & $-1$ & $-1$ & $-1$ & $-1$ & $-1$     \\
\hline
${\rm{Scalar}} $ & $\tilde{N}_{1}$ & $\tilde{N}_{\beta}$ & $D_{1}$ & $D^{\prime}_{1}$ & $D_{2}$ & $D^{\prime}_{2}$ & 
$\varphi$ & $\phi$ & $S$    \\
\hline 
$(B-L)$ & $5$ & $-4$ & $-4$ & $+4$ & $+5$ & $-5$ & $-10$ & $+8$ 
& $0$      \\
\hline 
$R-{\rm{Parity}}$ & $-1$ & $-1$ & $+1$ & $+1$ & $+1$ & $+1$ & $+1$ & $+1$ & $+1$     \\
\hline
\end{tabular}
\end{center}
\caption{Information about the $(B-L)$ quantum number and $R$-Parity of new fields of this model.}
\label{fermionblrm2}
\end{table}

Therefore, the lighest supersymmetric particle (LSP) 
is stable and a possible candidate to  Dark Matter, we will discuss about 
this subject latter in this article.

\section{The Lagrangian}
\label{sec:lagrangianm1}

With the superfields we can built a
supersymmetric invariant lagrangian. It has the following form
\begin{equation}
{\cal L}^{(B-L)} = {\cal L}_{SUSY} + {\cal L}_{soft}. \label{l1}
\end{equation}
Here, as usual, ${\cal L}_{SUSY}$ is the supersymmetric piece, while ${\cal L}_{soft}$ explicitly breaks SUSY. 
Below we will write ${\cal L}_{SUSY}$ in terms of the respective superfields. While in Sec.(\ref{sec:lagsofttermsm1}) 
we write ${\cal L}_{soft}$ in terms of the fields.

\subsection{The Supersymmetric terms}
\label{subsec:st}

The supersymmetric term can be divided as follows 
\begin{equation}
{\cal L}_{SUSY} = {\cal L}_{Quarks} + {\cal L}_{Gauge} + {\cal L}_{Lepton} +  {\cal L}_{Scalar},  
\label{l2}
\end{equation}
the terms ${\cal L}_{Quarks}, {\cal L}_{Gauge}$ are the same as in the MSSM and those terms in our notation is presented at~\cite{Rodriguez:2019,Rodriguez:2020,Rodriguez:2016esw}.

The term ${\cal L}_{Lepton}$ is given by
\begin{eqnarray}
{\cal L}_{Lepton}&=&{\cal L}^{cahrged}_{lepton}+ 
{\cal L}^{neutral}_{lepton},  
\label{l3}
\end{eqnarray} 
where
\begin{eqnarray}
{\cal L}^{cahrged}_{lepton}= \int d^{4}\theta\; \sum_{i=1}^{3}\left[\,
\hat{ \bar{L}}_{iL}e^{2[g\hat{W}+g^{\prime} \left( \frac{-1}{2} \right) \hat{b}^{\prime}]}\hat{L}_{iL}+ 
\hat{ \bar{E}}_{iR} e^{2[g^{\prime} \left( \frac{2}{2} \right) 
\hat{b}^{\prime}]}\hat{E}_{iR} \right].
\label{l3charg}
\end{eqnarray}
In the expressions above we have used $\hat{W}=T^{i}\hat{W}^{i}$ where $T^{i}=( \sigma^{i}/2)$ (with $i=1,2,3$) are the
generators of $SU(2)_{L}$ while $g^{\prime}$ is the gauge constant of $U(1)_{Y}$ see Table~\ref{table3}. The second term in Eq.(\ref{l3}) is 
written as  
\begin{eqnarray}
{\cal L}^{neutral}_{Lepton}= \int d^{4}\theta\; \left[\, 
\hat{ \bar{N}}_{1R}\hat{N}_{1R}+
\sum_{\beta =2}^{3}\hat{ \bar{N}}_{\beta R}\hat{N}_{\beta R}
\,\right].  
\label{l3neu}
\end{eqnarray}
Therefore, our right handed neutrinos, and the right-handed sneutrinos, they do not intercation with the usual gauge 
bosons then we will refere them as ``fully sterile" right handed 
neutrinos~\cite{Volkas:2001zb} and they\footnote{The right-handed neutrinos and the right-handed sneutrinos} can be Dark Matter 
candidate, as we will present at the end of Sec.(\ref{subsec:spotential}).

Finally, the scalar part in (\ref{l2}) is
\begin{eqnarray}
{\cal L}_{Scalar} &=& \int d^{4}\theta\;\left[\, 
\hat{ \bar{H}}_{1}e^{2[g\hat{W}+g{\prime} \left( \frac{1}{2} \right) \hat{b}^{\prime}]}\hat{H}_{1}+ 
\hat{ \bar{H}}_{2}e^{2[g\hat{W}+g^{\prime} \left( \frac{-1}{2}
\right) \hat{b}^{\prime}]}\hat{H}_{2} \,\
+ 
\hat{ \bar{D}}_{1}
e^{2[g\hat{W}+g^{\prime} \left( \frac{1}{2} \right) \hat{b}^{\prime}]}
\hat{D}_{1}\right.  \nonumber \\
&+& \left. 
\hat{ \bar{D^{\prime}}}_{1}
e^{2[g\hat{W}+g^{\prime} \left( - \frac{1}{2} \right) \hat{b}^{\prime}]}
\hat{D}^{\prime}_{1}+ 
\hat{ \bar{D}}_{2}
e^{2[g\hat{W}+g^{\prime} \left(  \frac{1}{2} \right) \hat{b}^{\prime}]}
\hat{D}_{2}+
\hat{ \bar{D}^{\prime}}_{2}
e^{2[g\hat{W}+g^{\prime} \left( - \frac{1}{2} \right) \hat{b}^{\prime}]}
\hat{D}^{\prime}_{2}\right.  \nonumber \\
&+& \left. 
\hat{ \bar{\varphi}}\hat{\varphi}+ 
\hat{ \bar{\phi}}\hat{\phi}\right]  +   
\left(\int d^2\theta\, W+ hc \right),
\label{l7}
\end{eqnarray}
where $W$ is the superpotential, which we discuss in the Sec.(\ref{subsec:spotential}). 

\subsection{The Superpotential}
\label{subsec:spotential}

In the non-supersymmetric version of this model, only the doublets $L_{iL},N_{1R},N_{\beta R},S$ 
of $SU(2)_{L}$ has $w$ as ${\cal Z}_{3}$ charges; the new scalars $D_{1,2}$ has 
$w^{-1}$ as ${\cal Z}_{3}$ charges and all the others fields $E_{iR}$ have this charge as identity~\cite{Machado:2017ksd}.

The superpotential of our model is given by
\begin{equation}
W=\frac{W_{2RC}}{2}+\frac{W_{3RC}}{3}+
\frac{W_{2RV}}{2}+ \frac{W_{3RV}}{3}+ hc,  
\label{sp1m1}
\end{equation}
where 
\begin{eqnarray}
W_{2RC}&=&\mu_{H}\left( \hat{H}_{1}\hat{H}_{2} \right) + 
\mu_{D_{1}}\left( \hat{D}_{1}
\hat{D}^{\prime}_{1}\right)+
\mu_{D_{2}}\left( \hat{D}_{2}
\hat{D}^{\prime}_{2}\right), 
\nonumber \\ 
W_{3RC}&=&
G^{d}\left( \hat{H}_{2}\hat{Q}_{iL}\right) \hat{d}_{jR}+
G^{u}\left( \hat{H}_{1}\hat{Q}_{iL}\right) \hat{u}_{jR}+
F_{ij}\left( \hat{H}_{2}\hat{L}_{iL}\right) \hat{E}_{jR}+
G_{i}(\hat{D}_{1}\hat{L}_{iL})
\hat{N}_{1R}
\nonumber \\ &+&
G_{i \beta}\left( \hat{D}_{2}\hat{L}_{iL}\right) \hat{N}_{\beta R} 
+
H_{11}\hat{\varphi}
\hat{N}_{1R}\hat{N}_{1R}+ 
H_{\alpha \beta}
\hat{\phi}
\hat{N}_{\alpha R}
\hat{N}_{\beta R}.
\label{wrc}
\end{eqnarray}
as usual we have defined 
\begin{equation}
\left( \hat{H}_{1}\hat{H}_{2} \right) \equiv \epsilon_{\alpha \beta} \hat{H}_{1}^{\alpha} \hat{H}_{2}^{\beta}.
\end{equation}
In general all the parameters $G^{d,u}$ and $F$ are, in principle, complex numbers and they are symmetric in $ij$ exchange and they are 
dimensionless parameters \cite{Baer:2006rs,dress}. 

Moreover, $G^{d}$ and $G^{u}$ can give account for the mixing between the quark current eigenstates 
as described by the Cabibbo-Kobayashi-Maskawa matrix (CKM matrix). In this model, we can also explain the mass hierarchy in the charged fermion masses as showed recently in 
\cite{cmmc,cmmc1}.
 
The couplings $H_{11}$ and $H_{\alpha \beta}$ will 
generate Majorana Mass terms to our right-handed 
neutrinos. Our superpotential is similar to Yukawa terms in the 
non-SUSY model and therefore three neutrinos get mass 
at tree level and we have three massless neutrinos, as we will show in 
Sec.(\ref{sec:neutrinomasses}).

The last two terms $W_{2RV},W_{3RV}$, 
defined at Eq.(\ref{sp1m1}), breaks the $(B-L)$ symmetry and they are written as
\begin{eqnarray}
W_{2RV}&=& \mu_{0i} \left( \hat{H}_{1}\hat{L}_{iL} \right) 
\nonumber \\
W_{3RV}&=&  
\lambda_{ijk}\left( \hat{L}_{iL}\hat{L}_{jL}\right) \hat{E}_{kR},
\label{wrpv}
\end{eqnarray}
the first term above allow the mixing between the Higgsinos with the usual leptons. The invariance under $SU(2)_{L}$ symmetry of the SM requires the antisymmetry of coupling $\lambda_{ijk}$ in $i,j$ and this parameter 
is, in principle, complex number. Therefore, we can expect to generate the phases 
of the Pontecorvo-Maki-Nakagawa-Sakata (PMNS) matrix, responsible for describe neutrino oscillation. 

In the MSSM model in $W_{RV}$ are include the terms $\lambda^{\prime}_{ijk}\left( \hat{L}_{iL}\hat{Q}_{jL}\right) \hat{d}_{kR}$ and $\lambda^{\prime\prime}_{ijk}\hat{u}_{iR}\hat{d}_{jR}\hat{d}_{kR}$
they are eliminated imposing the following 
${\cal Z}_{3}$ symmetry \cite{Haber:2000jh}
\begin{eqnarray}
\hat{L}, \hat{E}, \hat{N}, \hat{H}_{1,2},\hat{D}_{1,2},\hat{D}^{\prime}_{1,2}, \hat{\varphi} , \hat{\phi}&\rightarrow& \hat{L}, \hat{E}, \hat{N}, \hat{H}_{1,2},\hat{D}_{1,2},\hat{D}^{\prime}_{1,2}, \hat{\varphi} , \hat{\phi} \nonumber \\
\hat{Q}&\rightarrow&w \hat{Q}, 
\,\ 
\hat{u}, \hat{d}\rightarrow w^{-1} \hat{u}, w^{-1} \hat{d}, 
\end{eqnarray}
where $w=e^{(2 \imath \pi /3)}$, forbids the $B$-violating terms then we 
can avoid the proton decay, neutron-antineutron 
oscilations at tree level.

Therefore our symmetry are
\begin{eqnarray}
SU(3)_{C}\otimes SU(2)_{L}\otimes U(1)_{Y}\otimes 
U(1)_{B-L}\otimes {\cal Z}_{3},
\end{eqnarray}
where the symmetries $(B-L)$ and ${\cal Z}_{3}$ are global ones no local 
as the symmetry of the SM.

The interactions came at Eq.(\ref{wrpv}), which appears in the diagram to generate masses to left-handed neutrinos at 1-loop 
level, is given by  
\begin{eqnarray}
\lambda_{ijk}\left[ 
\tilde{\nu}^{i}_{L}\bar{l}^{k}_{R}l^{j}_{L}+ 
\tilde{l}^{j}_{L}\bar{l}^{k}_{R}\nu^{i}_{L}+
(\tilde{l}^{k}_{R})^{*}(\bar{\nu}^{i}_{L})^{c}l^{j}_{L}-
\left( i \longleftrightarrow j \right) + hc.
\right].
\label{neutralino-neutrino-sneutrinos} 
\end{eqnarray}
The coupling $\lambda$ can contribute to various (low-energy) 
process: charged current universality, bound on masses of 
$\nu_{e, \mu , \tau}$ and etc, for more details about this 
subject see~\cite{dress,Baer:2006rs}. 

The interactions in (\ref{wrpv}) induce neutrino masses, in 
the MSSM, at 1-loop can be written as \cite{dress}:
\begin{equation}
\delta m_{\nu_k}\propto (\lambda_{kii})^2
\frac{M_Sm^2_i}{\tilde{m}_i},
\label{numass1}
\end{equation}
where $m_i$ is the mass of exchanged fermion and the factor 
$m_{i}M_{S}$ comes from the left-right mixing of the sfermion.  
The interactions in (\ref{wrpv}) also induce flavour 
changing neutral currents, for instance, will generater the 
following decay:
\begin{equation}
\Gamma \left( 
\tilde{\nu}_{i} \to l^{+}_{j}l^{-}_{k}
\right) = \frac{1}{16 \pi}(\lambda_{ijk})^2m_{\tilde{\nu}_i},
\label{fcnc}
\end{equation}
where $m_{\tilde{\nu}_i}$ is the mass of sneutrinos and this 
decay violate lepton number conservation and it can generate Leptogenesis in not the same way as in the MSSM3RHN and we think it can be interesting to perform this analyses in this model. We have another 
interesting LSP decay, in the case of MSSM with $R$-parity conservation, given by
\begin{eqnarray}
\tilde{\chi}^{0}_{1} \to \bar{l}_{i}l_{j}\nu_{k},
\end{eqnarray}
and the decay from the lighest neutralino, the Dark Matter 
candidate at MSSM with $R$-parity conservation, produce 
missing energy in its decay. This decay can be observed if 
the appropriate $\lambda$ coupling satisfy the following 
relation
\begin{eqnarray}
| \lambda |>5 \times 10^{-7} \left(
\frac{m_{\tilde{l}}}{100 {\mbox GeV}} \right)^{2} \left(
\frac{100 {\mbox{GeV}}}{M_{LSP}}
\right)^{5/2},
\label{decaylspmssm}
\end{eqnarray}
where $m_{\tilde{l}}$ is the mass of charged slepton exchanged 
and $M_{LSP}$ is the mass of LSP, but the neutrinos will escape detection, 
leading to a missing transverse energy, but this decay will not be a problem in our case because the lighest neutralino is no more our Dark Matter candidate. In this model, the right-handed 
neutrinos $N_{iR}$ and the right-handed sneutrinos 
$\tilde{N}_{iR}$ 
are possible candidates to Dark Matter because they do not interct with the gauge bosons, see Eq.(\ref{l3neu}), and they interact only with the new scalars, their higgsinos, the leptons and sleptons, via the superpotential defined at 
Eq.(\ref{mssmrpc}), in a similar way as done at~\cite{Machado:2017ksd,Rodriguez:2020} but we will not consider this 
issue here. 

\subsection{Soft terms}
\label{sec:lagsofttermsm1}

Now we can write the soft terms as
\begin{eqnarray}
{\cal L}_{soft} &=& {\cal L}_{GMT}+{\cal L}_{SMT}+ 
{\cal L}_{INT}. \nonumber \\
\label{SoftSUSYm1}
\end{eqnarray}

The first term ${\cal L}_{GMT}$ give masses to all the gauginos in this model 
and it is written as 
\begin{eqnarray}
{\cal L}^{MSSM}_{GMT} &=&- \frac{1}{2}  \left[
\left(\,M_{3}\; \sum_{a=1}^{8} \lambda^{a}_{C} \lambda^{a}_{C}
+ M\; \sum_{i=1}^{3}\; \lambda^{i} \lambda^{i}
+ M^{\prime} \;   \lambda \lambda \,\right)
+ hc \right] \,\ .
\label{The Soft SUSY-Breaking Term prop 3}
\end{eqnarray}

The term ${\cal L}_{SMT}$, known as scalars mass term, is given by:
\begin{eqnarray}
{\cal L}_{SMT}&=&- \left( \sum_{i=1}^{3} \left[ 
M_{L}^{2}|\tilde{L}_{iL}|^{2}+
M^{2}_{l}|\tilde{E}_{iR}|^{2} \right] +
M^{2}_{N_{1}}|\tilde{N}_{1R}|^{2}+
\sum_{\beta =2}^{3}M^{2}_{N_{\beta}}|\tilde{N}_{\beta R}|^{2}  +
M^{2}_{H_{1}}|H_{1}|^{2}\right. \nonumber \\
&+& \left. 
M^{2}_{H_{2}}|H_{2}|^{2}+  
M^{2}_{D_{1}}|D_{1}|^{2}+
M^{2}_{D^{\prime}_{1}}|D^{\prime}_{1}|^{2}+
M^{2}_{D_{2}}|D_{2}|^{2}+M^{2}_{D^{\prime}_{2}}|D^{\prime}_{2}|^{2}+
M^{2}_{\varphi}|\varphi|^{2}\right. \nonumber \\
&+& \left. 
M^{2}_{\phi}|\phi|^{2} \right) + 
\left[ \beta_{H}\left( H_{1}H_{2} \right)+ \beta_{D_{1}}\left( D_{1}D^{\prime}_{1} \right)
+  
\beta_{D_{2}}\left( D_{2}D^{\prime}_{2} \right) + hc \right]. 
\label{lsoftint} 
\end{eqnarray}

The last term is given by
\begin{eqnarray}
{\cal L}_{Int}&=&\sum_{i,j,k=1}^{3}\left\{
A^{l}_{ij}G^{l}_{ij}\left( H_{2}\tilde{L}_{iL}\right) \tilde{E}_{jR}+
A^{\nu}_{i1}G_{i}\left( D^{\prime}_{1}\tilde{L}_{iL}\right) \tilde{N}_{1R}+
\sum_{\beta =2}^{3}A^{\nu}_{i \beta}P_{i \beta}\left( D^{\prime}_{2}\tilde{L}_{iL}\right) \tilde{N}_{\beta R}  
\right. \nonumber \\ &+& \left.
A^{M}_{11}H_{11}
\varphi \tilde{N}_{1R}\tilde{N}_{1R}+ 
\sum_{\alpha =2}^{3}\sum_{\beta =2}^{3}
A^{M}_{\alpha \beta}H_{\alpha \beta}
\phi \tilde{N}_{\alpha R}\tilde{N}_{\beta R}+
A^{d}_{ij}G^{d}_{ij}\left( \tilde{H}_{2}\tilde{Q}_{iL}\right) \tilde{d}_{jR}
\right. \nonumber \\ &+& \left.
A^{u}_{ij}G^{u}_{ij}\left( \tilde{H}_{1}\tilde{Q}_{iL}\right) \tilde{u}_{jR}+
A^{L}_{ijk}\left( \tilde{L}_{iL}\tilde{L}_{jL}\right) \tilde{E}_{kR}
+ hc  \right\}.
\label{lsoftint} 
\end{eqnarray}
The terms $A^{\nu},A^{M}$ can generate one physical CP violating phase at sneutrinos mass matrix \cite{Allahverdi:2006cx,Allahverdi:2009kr,Khalil:2009tm,Kajiyama:2009ae}.

The $A$-terms are known to play an important role in
Affleck-Dine baryogenesis~\cite{Montero:2016qpx,Rodriguez:2016esw}, as well as in the inflation
models based on supersymmetry~\cite{curvaton,AEGM,DHL}.

\section{Fermion masses}
\label{sec:fermionmasses}

We will present the masses of the usual fermions, it means the charged fermions, charginos, right-handed neutrinos, left-handed neutrinos and neutralinos at tree level.

\subsection{Charged Lepton masses}
In this case, it is possible to give mass to all
charged fermions. Denoting
\begin{equation}
\begin{array}{c}
\phi^{+}=(e^{+},\mu^{+},\tau^{+},- \imath 
\tilde{W}^{+}, \tilde{h}^{+}_{1})^{T},\\
\phi^{-}=(e^{-},\mu^{-},\tau^{-},- \imath \tilde{W}^{-},
\tilde{h}^{-}_{2})^{T},
\end{array}
\label{cbasismssm}
\end{equation}
where the winos, the supersymmetric $W^{\pm}$-boson partner, are defined as
\begin{equation}
\tilde{W}^{\pm} = \frac{1}{\sqrt{2}}\, \left( \tilde{W}^{1} \mp \imath \tilde{W}^{2} \right),
\label{wino2comp}
\end{equation}
all the fermionic fields are still Weyl spinors, we can define
$\Psi^{\pm}=(\phi^{+},\, \phi^{-})^{T}$, and the mass term   
$-(1/2)[\Psi^{\pm T}Y^{\pm}\Psi^{\pm}+hc]$ where 
$Y^{\pm}$ is the mass matrix given by: 
\begin{equation}
Y^{\pm}= \left( \begin{array}{cc}
0  & X^{T}\\
X  & 0
\end{array}
\right),
\label{ypmmssm}
\end{equation}
with
\begin{equation}
X= \left( \begin{array}{ccccc}
%%%%%%%%%linha 1 %%%%%%%%%%%%%%%
-F_{ee}v_{2}    & -F_{e\mu}v_{2}     & -F_{e\tau}v_{2}      &0 & 0  \\
%%%%%%%%%linha 2 %%%%%%%%%%%%%%%
-F_{e\mu}v_{2}  & -F_{\mu\mu}v_{2}   &  -F_{\mu\tau}v_{2}   & 0& 0 \\ 
%%%%%%%%%linha 3 %%%%%%%%%%%%%%%
-F_{e\tau}v_{2} & -F_{\mu\tau}v_{2}  & -F_{\tau\tau}v_{2}   & 0& 0\\ 
%%%%%%%%%linha 4 %%%%%%%%%%%%%%%
0            & 0               & 0      & 
M & \sqrt{2}M_{W}c_{\beta} \\
%%%%%%%%%linha 5 %%%%%%%%%%%%%%%
\mu_{0e} & \mu_{0\mu} & \mu_{0\tau}  & 
\sqrt{2}M_{W}s_{\beta}  & \mu_{H} 
\end{array}
\right),
\label{clmmmssm}
\end{equation} 
with $s_{\beta}=\sin\beta$, $c_{\beta}=\cos\beta$is 
defined the parameter  $\beta$ is defined at Eq.(\ref{betaparameter}), as we have presented at \cite{Montero:2001ch}.

The charginos mass matrix, is in similar way as done in the MSSM, diagonalized by two unitaries matrix $U$ and $V$ \cite{Rodriguez:2019,Rodriguez:2020,dress,Baer:2006rs,
ait1,Rodriguez:2016esw}
\begin{eqnarray}
\tilde{\chi}^{+}_{i}&=& \sum_{j=1}^{5} U_{ij} \phi^{+}_{j}, \nonumber \\
\tilde{\chi}^{-}_{i}&=&
\sum_{j=1}^{5} V_{ij} \phi^{-}_{j}.
\end{eqnarray}

In the Eq.(\ref{clmmmssm}), we have two sector. The first sector is very light and we will associate it with the usual charged lepton. The second sector has heavier masses than the first one, the usual charginos at MSSM, for more details see \cite{Montero:2001ch}.

The masses of the quarks are the same as in the MSSM \cite{Rodriguez:2019,Rodriguez:2020,dress,Baer:2006rs,
ait1,Rodriguez:2016esw}.

\subsection{Mixing Neutrinos with Gauginos at Tree level}
\label{sec:neutrinomasses}

Our right-handed neutrinos get Majorana mass term, due the 
following terms
\begin{eqnarray}
H_{11}\varphi N_{1R}N_{1R}+
H_{\alpha \beta}\phi N_{\alpha R}N_{\beta R},
\end{eqnarray}
in our superpotential defined at Eq.(\ref{wrc}). As the 
new scalars $D_{1},D_{2}$ are innert due $(B-L)$ symmetry, 
see Eq.(\ref{vevextradoublets}), the Yukawa couplings
\begin{eqnarray}
G_{i}\left( D_{1}L_{iL}\right) N_{1R}+
P_{i \beta}\left( D_{2}L_{iL}\right)
N_{\beta R}+hc,
\label{yukawacoupleftneutrinos}
\end{eqnarray}
coming from our superpotential, defined at Eq.(\ref{wrc}), 
does not generate mixing between left-handed neutrinos with the 
right-handed neutrinos, in similar way as happen in the non-SUSY 
model \cite{Machado:2017ksd}.

Defining the basis $\Psi^{0}=(N_{1},N_{2},N_{3},\nu_{e},
\nu_{\mu},\nu_{\tau},- \imath \tilde{W}^{3},- \imath \tilde{b}^{\prime},\tilde{h}^{0}_{2},
\tilde{h}^{0}_{1})^{T}$, the mass term is 
$-(1/2)[\Psi^{0T}Y^{0}\Psi^{0}+hc]$, where 
$Y^{0}$ is the mass matrix given by
\begin{equation}
Y^{0}= \left(
\begin{array}{ccc}
M_{M} & 0_{3 \times 3} & 0_{4 \times 3} \\ 
0_{3 \times 3} & 0_{3 \times 3} & m_{4 \times 3} \\
0_{3 \times 4} & \left( m_{4 \times 3} \right)^{T} & M^{neu}_{MSSM}
\end{array}
\right).
\label{majorananeutrino}
\end{equation}

We see, we have two sector. The first ones is given by the right-handed neutrinos and the second one are, similar, to the neutralinos from the MSSM.

In order to write $M_{M}$, first we  define $\Gamma$   through  the ratio
\begin{eqnarray}
\tan \Gamma &=& \frac{u_{1}}{u_{2}}, 
\label{angles1}
\end{eqnarray}
and $M_{M}$ reads
\begin{equation}
M_{M}= \frac{2u_{2}}{\sqrt{2}}\left(
\begin{array}{ccc}
f^{M}_{11} & 0 & 0 \\
0 & f^{M}_{22} \tan \Gamma & f^{M}_{23} \tan \Gamma \\
0 & f^{M}_{32}\tan \Gamma & f^{M}_{33} \tan \Gamma
\end{array}
\right).
\label{majorananeutrino}
\end{equation}
This sector is the same as the right-handed neutrinos at non-SUSY model presented at \cite{Machado:2017ksd}.

In the neutralinos sector, we can define, the following mass matrices
\begin{eqnarray}
m_{4 \times 3}&=& \left(
\begin{array}{cccc}
0  & 0 & 0& -\mu_{0e}\\
0  & 0 & 0& -\mu_{0\mu} \\
0  & 0 & 0& -\mu_{0\tau}
\end{array}
\right), \nonumber \\
M^{neu}_{MSSM}&=& \left(
\begin{array}{cccc}
M & 0 &  M_{Z}\,s_{\beta} c_{W}& 
-M_{Z}\,c_{\beta} c_{W}\\
0 & M^{\prime} & M_{Z}\,s_{\beta} s_{W}& 
-M_{Z}\,c_{\beta} s_{W} \\
M_{Z}\,s_{\beta} c_{W}& M_{Z}\,s_{\beta} s_{W}
& 0& \mu_{H} \\
-M_{Z}\,c_{\beta} c_{W}& -M_{Z}\,c_{\beta} s_{W}& 
\mu_{H}& 0 
\end{array}
\right),
\label{majorananeutrino}
\end{eqnarray}
with $s_{W}=\sin\theta_{W}$, $c_{W}=\cos\theta_{W}$ 
where $\theta_W$ is the weak mixing. This mass matrix is 
diagonalized by an unitary $7 \times 7$ matrix $N$, defined as \cite{Rodriguez:2019,Rodriguez:2020,dress,Baer:2006rs,
ait1,Rodriguez:2016esw}
\begin{eqnarray}
\tilde{\chi}^{0}_{i}&=& \sum_{j=1}^{7} N_{ij} \phi^{0}_{j},
\end{eqnarray}
where we have defined $\phi^{0}=(\nu_{e},
\nu_{\mu},\nu_{\tau},- \imath \tilde{W}^{3},- \imath \tilde{b}^{\prime},\tilde{h}^{0}_{2},
\tilde{h}^{0}_{1})^{T}$.

We obtain
besides the two massless neutrinos a massive one with, 
we can choose the parameters in such way that
$m_{\nu_{H}}=5\times10^{-2}$ eV, to explain the 
atmospheric neutrino, in similar way as done at \cite{Montero:2001ch}.

The mixing PMNS matrix is defined as
\begin{equation}
\left(
\begin{array}{c}
\nu_{e} \\
\nu_{\mu} \\
\nu_{\tau}
\end{array}
\right)_{L}=
U_{PMNS}
\left(
\begin{array}{c}
\nu^{\prime}_{e} \\
\nu{\prime}_{0} \\
\nu_{H}
\end{array}
\right)_{L}
\label{pmnsdef}
\end{equation}
when we have one massive neutrino massive $\nu_{H}$ 
and two massles $\nu^{\prime}_{e}$ and 
$\nu{\prime}_{0}$  we can define $U_{PMNS}$ in the 
following way \cite{King:1998jw}
\begin{equation}
U_{PMNS}=
\left(
\begin{array}{ccc}
%%%% First Line
1 & \left( \frac{c_{23}}{s_{23}} \right) \theta_{1} 
& \theta_{1} \\
%%% Second Line
- \left( \frac{\theta_{1}}{s_{23}} \right) & 
c_{23} \left( 1+ \sin^{2}\theta_{23} 
\frac{m_{\nu_{\tau}}}{m_{\nu_{H}}} \right) & s_{23} \left( 1- \cos^{2}\theta_{23} 
\frac{m_{\nu_{\tau}}}{m_{\nu_{H}}} \right) \\
%%% Third Line
0 & -s_{23}\left( 1- \cos^{2}\theta_{23} 
\frac{m_{\nu_{\tau}}}{m_{\nu_{H}}} \right) & c_{23}\left( 1+ \sin^{2}\theta_{23} 
\frac{m_{\nu_{\tau}}}{m_{\nu_{H}}} \right)
\end{array}
\right),
\label{pmnsparame}
\end{equation}
where $s_{23}= \sin \theta_{23}$, 
$c_{23}= \cos \theta_{23}$ and $(m_{\nu_{\tau}}/2)\sim 3 \times 10^{-3}$ eV is the mass of neutrino of tau and it is an order of magnitude smaller then $m_{\nu_{H}}$.

It is natural, we choose $\mu_{0e}\ll \mu_{0 \mu}\approx 
\mu_{0 \tau}$, because they break softky our 
$B-L$ global symmetry. In this case, we can define a 
small angle $\theta_{1}$ defined as
\begin{equation}
\theta_{1}\approx 
\frac{\mu_{0e}}
{\sqrt{|\mu_{0e}|^{2}+|\mu_{0 \mu}|^{2}+|\mu_{0 \tau}|^{2}}},
\end{equation}
we can solve the solar neutrino problem imposing the following constraints 
$\sin^{2}\theta_{1}\sim 10^{-2}$. There is a big mixing angle $\theta_{23}$ in the following 
way
\begin{eqnarray}
\sin \theta_{23}\approx 
\frac{\mu_{0 \mu}}
{\sqrt{|\mu_{0e}|^{2}+|\mu_{0 \mu}|^{2}+|\mu_{0 \tau}|^{2}}}, \nonumber \\
\cos \theta_{23}\approx 
\frac{\mu_{0 \tau}}
{\sqrt{|\mu_{0e}|^{2}+|\mu_{0 \mu}|^{2}+|\mu_{0 \tau}|^{2}}}.
\end{eqnarray}
To explain the atmospheric neutrinos data we can 
choose $\theta_{23}\sim ( \pi /4)$.

Therefore only the right handed neutrinos get masses at tree level as 
happen at non-SUSY model presented at \cite{Machado:2017ksd}. Therefore, 
the two left-handed neutrinos are massless and they will get masses at 
1-loop level, as we will discuss at Sec.(\ref{sec:1loopmechanism}).

\section{Scalar Potential}
\label{sec:scalarpotential}

The Higgs potential of our model has the following form
\begin{equation}
V_{scalar}=V_{F}+V_{soft}+V_{D}.
\label{vfdsoft}
\end{equation}
where, we are only writing the terms of scalars in doublets because we 
want to compare our scalar potential with ones presented at non-SUSY 
model presented at \cite{Machado:2017ksd} 
\begin{eqnarray}
%%%%%% Primeira linha (F_{H_{1}},F_{H_{2}})
V_{F}&=& \frac{\mu^{2}_{H}}{4}\left( |H_{1}|^{2}+|H_{2}|^{2} \right)+
%%%%%% Segunda linha (F_{D_{1}},F_{D^{\prime}_{1}})
\frac{\mu^{2}_{D_{1}}}{4}\left( |D_{1}|^{2}+|D^{\prime}_{1}|^{2} \right)+
%%%%%% Terceira linha (F_{D_{2}},F_{D^{\prime}_{2}})
\frac{\mu^{2}_{D_{2}}}{4}\left( |D_{2}|^{2}+|D^{\prime}_{2}|^{2} \right), 
\nonumber \\
V_{soft}&=& M^{2}_{H_{1}}|H_{1}|^{2}+ M^{2}_{H_{2}}|H_{2}|^{2} +  
M^{2}_{D_{1}}|D_{1}|^{2}+M^{2}_{D^{\prime}_{1}}|D^{\prime}_{1}|^{2}+ 
M^{2}_{D_{2}}|D_{2}|^{2}+ 
M^{2}_{D^{\prime}_{2}}|D^{\prime}_{2}|^{2}
\nonumber \\ &-&
\left[ \beta_{H}\left( H_{1}H_{2} \right)+
\beta_{D_{1}}\left( D_{1}D^{\prime}_{1} \right)  +  
\beta_{D_{2}}\left( D_{2}D^{\prime}_{2} \right) + hc \right], \nonumber \\
V_{D}&=&
\left( \frac{(g^{\prime})^{2}}{8}+ \frac{g^{2}}{8} \right)\left( |H_{1}^{\dagger}H_{1}|^{2}+|H_{2}^{\dagger}H_{2}|^{2} \right) +
\left(\frac{(g^{\prime})^{2}}{8}+ \frac{g^{2}}{8} \right) 
\left( |D_{1}^{\dagger}D_{1}|^{2}+|D^{\dagger}_{1}D^{\prime}_{1}|^{2} \right) 
 \nonumber \\ &+& 
\left(\frac{(g^{\prime})^{2}}{8}+  \frac{g^{2}}{8} \right) 
\left( |D_{2}^{\dagger}D_{2}|^{2}+|D^{\dagger}_{2}D^{\prime}_{2}|^{2} \right) 
\nonumber \\ &+& 
\frac{g^{2}}{4}\left(
|H_{1}^{\dagger}H_{2}|^{2}+|H_{1}^{\dagger}D_{1}|^{2}+|H_{1}^{\dagger}D^{\prime}_{1}|^{2}+
|H_{1}^{\dagger}D_{2}|^{2}+ |H_{1}^{\dagger}D^{\prime}_{2}|^{2}+
|H_{2}^{\dagger}D_{1}|^{2}+|H_{2}^{\dagger}D^{\prime}_{1}|^{2}
\right. \nonumber \\ &+& \left. \hspace{-4mm}
|H_{2}^{\dagger}D_{2}|^{2}+|H_{2}^{\dagger}D^{\prime}_{2}|^{2}
+
|D_{1}^{\dagger}D^{\prime}_{1}|^{2}+|D_{1}^{\dagger}D_{2}|^{2}+
|D_{1}^{\dagger}D^{\prime}_{2}|^{2}+
%\right. \nonumber \\ &+& \left. 
|D^{\prime \dagger}_{1}D_{2}|^{2}
+ |D^{\prime \dagger}_{1}D^{\prime}_{2}|^{2}+|D^{\dagger}_{2}D^{\prime}_{2}|^{2} 
\right)  
\nonumber \\
&-& \left( \frac{(g^{\prime})^{2}}{4}+ \frac{g^{2}}{8} \right) \left(
|H_{1}|^{2}|H_{2}|^{2}+|H_{1}|^{2}|D^{\prime}_{1}|^{2}
+|H_{1}|^{2}|D^{\prime}_{2}|^{2}
+|H_{2}|^{2}|D_{1}|^{2}
+|H_{2}|^{2}|D_{2}|^{2}
\right.  \nonumber \\ &+& \left. 
|D_{1}|^{2}|D^{\prime}_{1}|^{2}+
|D_{1}|^{2}|D^{\prime}_{2}|^{2}
+|D^{\prime}_{1}|^{2}|D_{2}|^{2}+
|D_{2}|^{2}|D^{\prime}_{2}|^{2}
\right) 
\nonumber \\ &+& 
\left( \frac{(g^{\prime})^{2}}{4}- \frac{g^{2}}{8} \right) \left(
|H_{1}|^{2}|D_{1}|^{2}+|H_{1}|^{2}|D_{2}|^{2}+
|H_{2}|^{2}|D^{\prime}_{1}|^{2}+|H_{2}|^{2}|D^{\prime}_{2}|^{2}+
|D_{1}|^{2}|D_{2}|^{2}
\right.  \nonumber \\ &+& \left.
|D^{\prime}_{1}|^{2}|D^{\prime}_{2}|^{2}
\right),
\end{eqnarray}
and we see we can these scalar potential are very similar to ones defined in the non-SUSY model presented at \cite{Machado:2017ksd}. As it is done at MSSM when compare with the Two Higgs Double Model, we can write the following relations
\begin{eqnarray}
\mu^{2}_{SM}&=&\frac{\mu^{2}_{H}}{4}+M^{2}_{H_{1}}, \nonumber \\ 
\mu^{2}_{d1}&=&\frac{\mu^{2}_{D_{1}}}{4}+M^{2}_{D^{\prime}_{1}}, \nonumber \\
\mu^{2}_{d2}&=&\frac{\mu^{2}_{D_{2}}}{4}+M^{2}_{D_{2}}, \nonumber \\
\lambda_{1}&=&\lambda_{2}=
\lambda_{3}= \left(\frac{(g^{\prime})^{2}}{8}+ \frac{g^{2}}{8} \right), 
\nonumber \\
\lambda_{4}&=&\lambda_{5}=
\lambda_{6}= \left(\frac{(g^{\prime})^{2}}{8}- \frac{g^{2}}{8} \right), 
\nonumber \\
\lambda_{7}&=&\lambda_{8}= 
\frac{g^{2}}{4},
\label{lambdanonosusy}
\end{eqnarray}
we can conclude we can not get the terms $\mu^{2}_{12},\lambda_{9,10}$ 
defined in the non-SUSY model and 
therefore we can not have scotogenic 
mechanism, but we will still hace the 
1-loop correction drawn at their Figure 
1, as we will discuss in the next section.

In order to generate the term $\mu^{2}_{12}$, as it is quadratic in the 
usual scalars we would have to introduce term like
\begin{eqnarray}
\bar{D}_{1}D_{2},
\end{eqnarray}
in our superpotential defined at Eq.(\ref{sp1m1}), but it is not chiral 
and therefore we can not introduce it. As we can not introduce this term 
we can not have
\begin{eqnarray}
\mu^{2}_{12}D^{\dagger}_{1}D_{2},
\end{eqnarray}
at soft terms, see Eq.(\ref{lsoftint}), and therefore due SUSY 
algebra we can not generate this term in this supersymmetric model.

We can not generate the terms proportional to $\lambda_{9,10}$and it is due 
the fact to generate them we need to introduce term like 
\begin{equation}
\hat{ \bar{H}}e^{2[g \hat{W}]}\hat{D}_{1,2},
\end{equation}
this term would give the following contribution to $D$-term
\begin{equation}
gD^{i}H^{\dagger}T^{i}D_{1,2}
\end{equation}
and
\begin{equation}
(H^{\dagger}\sigma^{i}D_{1,2})(H^{\dagger}\sigma^{i}D_{1,2})=
2(H^{\dagger}D_{1,2})(H^{\dagger}D_{1,2})-(H^{\dagger}D_{1,2})(H^{\dagger}D_{1,2})=(H^{\dagger}D_{1,2})^{2},
\end{equation}
but this term is not invariant under our global $(B-L)$ symmetry.

\section{$1$-loop mechanism to generate masses to the left-handed neutrinos}
\label{sec:1loopmechanism}

The left-handed neutrinos get masses at 1-loop level due the following interaction defined at Eq.(\ref{yukawacoupleftneutrinos}). As 
we have the couplings $\lambda_{4}$ and $\lambda_{5}$, see 
Eq.(\ref{lambdanonosusy}), we 
can get the first diagram to 
generate neutrinos masses at 1-loop level drawn at Figure 1 presented at~\cite{Machado:2017ksd}. 

\begin{figure}[ht]
\begin{center}
\vglue -0.009cm
\mbox{\epsfig{file=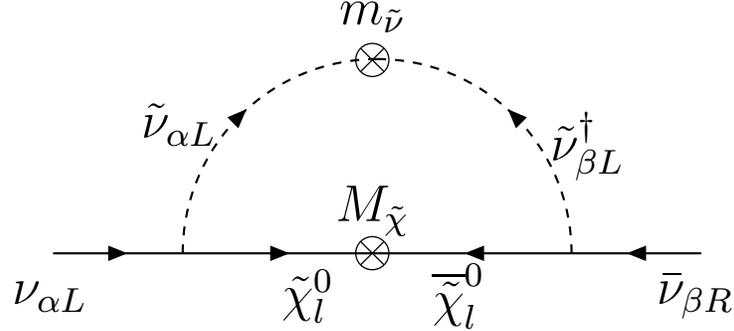,width=0.7\textwidth,angle=0}}       
\end{center}
\caption{The one loop correction to the masses of 
$m_{\nu_{\alpha}}$ inclunding (gauginos)neutralino-neutrino-sneutrino 
vertices, where $\alpha =e, \mu , \tau$, this vertices came from 
Eq.(\ref{verticeinferior}), and is given by $(- \imath /2)(gZ^{\ast}_{l1}\sin \theta_{W}+
g^{\prime}Z^{\ast}_{l2}\cos \theta_{W})$.}
\label{fig1}
\end{figure}

\begin{figure}[ht]
\begin{center}
\vglue -0.009cm
\mbox{\epsfig{file=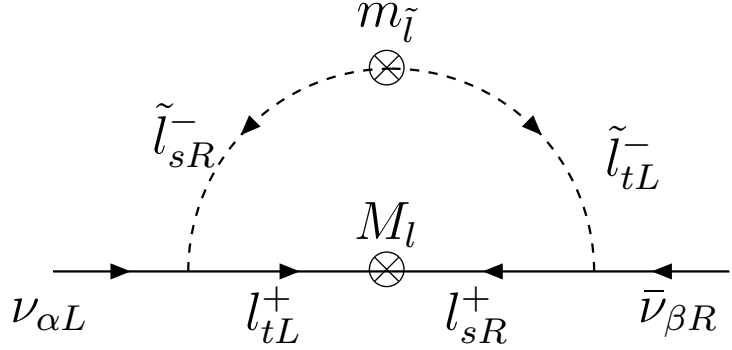,width=0.7\textwidth,angle=0}}       
\end{center}
\caption{The one loop correction to the masses of $m_{\nu_{\alpha}}$ inclunding charged lepton-neutrino-charged slepton vertices, this vertices came from $W_{RV}$ see Eq.(\ref{wrpv}), and the left vertices 
is proportional to $\lambda_{\alpha st}$ and the right ones is 
$\lambda_{\beta st}$.}
\label{fig2}
\end{figure}

In this supersymmetric case, we can get two more contribution to generate neutrinos masses at 1-loop level. The first contribution  is due the sneutrinos masses and it is 
drawn at Figure \ref{fig1}. The coupling $\lambda_{ijk}$ defined 
at Eq.(\ref{wrpv}) generate the  1-loop diagrams drawn at 
Figure \ref{fig2}.

The quartic interactions between left-handed neutrino-left handed 
sneutrino-usual scalars  is the same as appear at MSSM therefore it is given by \cite{dress}
\begin{eqnarray}
\imath \left( D[ \phi , \phi^{\prime}, \tilde{f}^{\prime}, 
\tilde{f}] + D[ \phi^{\prime}, \phi , \tilde{f}^{\prime}, 
\tilde{f}] \right),
\label{leftneutrinos-leftsneutrinos-neutralinos}
\end{eqnarray}
for $\phi = \phi^{\prime}$, only one of the two contributions is 
non zero, whereas for $\phi \neq \phi^{\prime}$ both contributions are equal. The coefficients of various quartic 
interaction are:
\begin{eqnarray}
D[H,H, \tilde{\nu}_{i}, \tilde{\nu_{j}}]&=&-d_{g}[ \tilde{\nu_{i}}] 
c_{2 \alpha}\delta_{ij}, \,\
D[H,h, \tilde{\nu}_{i}, \tilde{\nu_{j}}]=2d_{g}[ \tilde{\nu_{i}}] 
s_{2 \alpha}\delta_{ij} , \nonumber \\
D[h,h, \tilde{\nu}_{i}, \tilde{\nu_{j}}]&=&d_{g}[ \tilde{\nu_{i}}] 
c_{2 \alpha}\delta_{ij}, \,\
D[A,A, \tilde{\nu}_{i}, \tilde{\nu_{j}}]=d_{g}[ \tilde{\nu_{i}}] 
c_{2 \beta}\delta_{ij}, \nonumber \\
D[H,H, \tilde{l}_{s}, \tilde{l_{t}}]&=&-
d_{Y}[ \tilde{l}_{s}, \tilde{l_{t}}]c^{2}_{ \alpha}-
d_{g}[ \tilde{l}_{s}, \tilde{l_{t}}]c_{2 \alpha}, 
\,\
D[H,h, \tilde{l}_{s}, \tilde{l_{t}}]=
d_{Y}[ \tilde{l}_{s}, \tilde{l_{t}}]s_{2 \alpha}+2
d_{g}[ \tilde{l}_{s}, \tilde{l_{t}}]s_{2 \alpha}, \nonumber \\
D[h,h, \tilde{l}_{s}, \tilde{l_{t}}]&=&-
d_{Y}[ \tilde{l}_{s}, \tilde{l_{t}}]s^{2}_{ \alpha}+
d_{g}[ \tilde{l}_{s}, \tilde{l_{t}}]c_{2 \alpha}, \,\
D[A,A, \tilde{l}_{s}, \tilde{l_{t}}]=-
d_{Y}[ \tilde{l}_{s}, \tilde{l_{t}}]s^{2}_{ \beta}+
d_{g}[ \tilde{l}_{s}, \tilde{l_{t}}]c_{2 \beta}
, \nonumber \\
\end{eqnarray}
where the rotations angle $\alpha$ and $\beta$ seen to obey 
the relations \cite{dress,Baer:2006rs}
\begin{eqnarray}
\sin (2 \alpha)&=&- 
\frac{M^{2}_{H^{0}}+M^{2}_{h^{0}}}{M^{2}_{H^{0}}-M^{2}_{h^{0}}}\sin (2 \beta), 
\,\
\cos (2 \alpha)=-
\frac{M^{2}_{A^{0}}-M^{2}_{Z}}{M^{2}_{H^{0}}-M^{2}_{h^{0}}}\cos (2 \beta), 
\nonumber \\
\tan (2 \alpha)&=& 
\frac{M^{2}_{H^{0}}+M^{2}_{h^{0}}}{M^{2}_{A^{0}}-M^{2}_{Z}}
\tan (2 \beta),
\end{eqnarray}
where $h$ is the lighest CP-even Higgs, $H$ the heavy CP-even 
Higgs and $A$ is the pseudo-scalar of MSSM. For any angle $\zeta$, we use $s_{\zeta},c_{\zeta},t_{\zeta}$ 
to mean $\sin \zeta$, $\cos \zeta$ and $\tan \zeta$ respectively.

We have, also, defined
\begin{eqnarray}
d_{g}[ \tilde{\nu}_{i}]&=& \frac{g^{2}}{8} \left( 1+t^{2}_{W} \right), \,\
d_{g}[ \tilde{l}_{s}, \tilde{l_{t}}]=
\frac{g^{2}}{4M^{2}_{W}c^{2}_{\beta}}m^{2}_{\tilde{f}}\left[
\cos \left( \theta_{\tilde{s}}- \theta_{\tilde{t}} \right) 
\right], \nonumber \\
d_{Y}[ \tilde{l}_{s}, \tilde{l_{t}}]&=&- \frac{g^{2}}{8} \left[
2t^{2}_{W} \sin \theta_{\tilde{s}} \sin \theta_{\tilde{t}}
+
\cos \theta_{\tilde{s}} \cos \theta_{\tilde{t}} 
\left( 1-t^{2}_{W} \right) 
\right], \nonumber \\
\end{eqnarray}
where $\theta_{\tilde{f}}$ is the mixing angle at charged 
slepton, $m^{2}_{\tilde{f}}$ their mass, while the symbols 
to Weinberg angle $\theta_{W}$ are $s_{W},c_{W},t_{W}$.

In the neutralinos sector, the mass eigenstates, see 
\cite{Rodriguez:2019}, are defined as
\begin{equation}
\tilde{\chi}^{0}_{i} = Z_{ij}\,\psi^{0}_{j}, \hspace{6mm} i,j=1\ldots 4,
\label{mixingatsneutralinos}
\end{equation}
where  
\begin{eqnarray}
\psi^{0}_{l}=
\left(
\begin{array}{cccc}
- \imath \tilde{W}^{3},- \imath \tilde{b}^{\prime},\tilde{H}_{1}, \tilde{H}_{2}
\end{array}
\right)^{T}.
\label{neutralinossymmetryeigenstates}
\end{eqnarray} 
Therefore, the light sector of neutralinos of this model is exactly the same as in the MSSM and we can say that $M_{\tilde{\chi}^{0}_{1}}\sim {\cal O}(100)$ MeV as we hope at MSSM.

The mass matrix to the sneutrinos when we have three right-handed 
neutrinos can be write as \cite{Rodriguez:2020}
\begin{equation}
M^{2}_{sneutrinos}= \left(
\begin{array}{cc}  
m^{2}_{\tilde{\nu}_{eL}}  & A^{\nu}f^{\nu}v_{2} \\
A^{\nu}f^{\nu}v_{2} & m^{2}_{\tilde{\nu}_{eR}} 
\end{array}\right),
\end{equation}
but $m_{\nu}=A^{\nu}f^{\nu}v_{2}$ is the neutrinos masses at tree level 
and $\tilde{\nu}_{e},\tilde{\nu}_{\nu}$ and $\tilde{\nu}_{\tau}$ are 
already mass eigenstates, as first approximation, and we will represent 
them as $\tilde{\nu}_{\alpha}$. Therefore the 
left handed neutrino $\nu_{\alpha}$ can couple only with its 
left-handed sneutrnios, this vertice came from 
${\cal L}^{cahrged}_{lepton}$ defined at 
Eq.(\ref{l3charg}).

In this model, the left-handed neutrinos do not get masses at tree level. 
and the coupling at soft superssymetric terms
\begin{eqnarray}
A^{\nu}_{i1}G^{\nu}_{i}\left( D_{1}\tilde{L}_{iL}\right) \tilde{N}_{1R}+
A^{\nu}_{i \beta}F^{\nu}_{i \beta}\left( D_{2}\tilde{L}_{iL}\right)
\tilde{N}_{\beta R}+hc,
\label{mixiningsneutrinosector}
\end{eqnarray}
will induce an effective mixing between our left-handed sneutrinos with 
the right-handed sneutrinos and it is the responsible for the mixing in 
this sector.

The interactions between left-handed neutrino-left handed 
sneutrino-neutralino, came from 
${\cal L}^{cahrged}_{lepton}$ see Eqs.(\ref{l3charg},\ref{leftneutrinos-leftsneutrinos-neutralinos}),thata appear in our 1-loop correction 
drawn at Figure \ref{fig1}, is written as
\begin{eqnarray}
\frac{- \imath g}{\sqrt{2}}
\nu_{\alpha}\tilde{W}^{3}\tilde{\nu}^{\dagger}_{\alpha}-
\frac{\imath g^{\prime}}{\sqrt{2}}\nu_{\alpha}
\tilde{b}^{\prime}\tilde{\nu}^{\dagger}_{\alpha}+hc,
\end{eqnarray}
now using 
Eqs.(\ref{mixingatsneutralinos},\ref{neutralinossymmetryeigenstates}) 
at equation above we get
\begin{eqnarray}
\frac{- \imath gZ^{\ast}_{l1}}{\sqrt{2}} \sin \theta_{W}
\nu_{\alpha}\tilde{\chi}^{0}_{l}\tilde{\nu}^{\dagger}_{\alpha}-
\frac{\imath g^{\prime}Z^{\ast}_{l2}}{\sqrt{2}} \cos \theta_{W}
\nu_{\alpha}\tilde{\chi}^{0}_{l}\tilde{\nu}^{\dagger}_{\alpha}+hc,
\label{verticeinferior}
\end{eqnarray}
this is the vertices give contribution to generate masses to 
left-handed neutrinos 
at 1-loop level, its diagram is drawn at 
Figure \ref{fig1}
\footnote{We get Majorana Mass terms to our neutrinos}, those 
kind of mechanism were discussed at~\cite{Smirnov:2004hs,GH}.

Our right-handed sneutrinos $\tilde{N}_{iR}$ are possibles candidates 
to be the Dark Matter together with the right handed neutrinos $N_{iR}$.

The coupling $\lambda_{ijk}$ defined at Eq.(\ref{wrpv}) and 
the mixing between the sleptons is defined as
\begin{equation}
\left(
\begin{array}{c}
\tilde{l}_{1} \\
\tilde{l}_{2}
\end{array}
\right) =
\left(
\begin{array}{cc}
\cos \theta_{\tilde{l}} & \sin \theta_{\tilde{l}}  \\
- \sin \theta_{\tilde{l}} & \cos \theta_{\tilde{l}}
\end{array}
\right)
\left(
\begin{array}{c}
\tilde{l}_{L} \\
\tilde{l}_{R}
\end{array}
\right).
\label{mixingchargedleptons}
\end{equation} 
Therefore, the charged sleptons sector of this model is exactly the same as in the MSSM. The charged sleptons  generate the  1-loop diagrams drawn at Figure \ref{fig2}.

The contribution to neutrinos masses at 1-loop correction is   we get the following one loop correction to the left-handed 
neutrinos masses expressions to the $\nu_{e}$
\begin{eqnarray}
\left( m_{\nu} \right)_{ij}&=&
\frac{G_{i}G_{j}M_{1}}{32 \pi^{2}} \left[
\frac{m^{2}_{R1}}{m^{2}_{R1}-M^{2}_{1}} \ln\left( \frac{m^{2}_{R1}}{M^{2}_{1}}\right)-
\frac{m^{2}_{I1}}{m^{2}_{I1}-M^{2}_{1}} \ln\left( \frac{m^{2}_{I1}}{M^{2}_{1}}\right)
\right]
\nonumber \\ &+&
\frac{F_{ik}F_{jk}M_{K}}{32 \pi^{2}} \left[
\frac{m^{2}_{R2}}{m^{2}_{R2}-M^{2}_{k}} \ln\left( \frac{m^{2}_{R2}}{M^{2}_{k}}\right)-
\frac{m^{2}_{I2}}{m^{2}_{I2}-M^{2}_{k}} \ln\left( \frac{m^{2}_{I2}}{M^{2}_{k}}\right)
\right]
\nonumber \\ &+&\hspace{-4mm}
%\sum_{l=1}^{4}
\frac{
\left( gZ^{\ast}_{l1}\sin \theta_{W}+
g^{\prime}Z^{\ast}_{l2}\cos \theta_{W}\right)^{2} M_{\tilde{\chi}^{0}_{l}}}{64 \pi^{2}}
\left[ 
\frac{m^{2}_{\tilde{\nu_{i}}}}{m^{2}_{\tilde{\nu_{i}}}-M^{2}_{\tilde{\chi}^{0}_{l}}}
\ln\left( \frac{m^{2}_{\tilde{\nu_{i}}}}{M^{2}_{\tilde{\chi}^{0}_{l}}}  \right)-
\frac{m^{2}_{\tilde{\nu_{j}}}}{m^{2}_{\tilde{\nu_{j}}}-M^{2}_{\tilde{\chi}}}
\ln\left( \frac{m^{2}_{\tilde{\nu_{j}}}}{M^{2}_{\tilde{\chi}^{0}_{l}}}  \right)  \right]
\nonumber \\ &+& \hspace{-4mm}
%\sum_{i=1}^{3}
\frac{\lambda_{its}\lambda_{jst}}{16 \pi^{2}}
\sin^{2}\left( \theta_{\tilde{s}}+ \theta_{\tilde{t}}\right)
\left[  
\frac{m^{2}_{\tilde{s}}M_{s}}{m^{2}_{\tilde{s}}-M^{2}_{s}}
\ln\left( \frac{m^{2}_{\tilde{s}}}{M^{2}_{s}}  \right)- 
\frac{m^{2}_{\tilde{t}}M_{t}}{m^{2}_{\tilde{t}}-m^{2}_{t}}
\ln\left( \frac{m^{2}_{\tilde{t}}}{M^{2}_{t}}  \right)  \right], 
\end{eqnarray}
where $Z_{lk}$ is the mixing in the neutralino sector, see 
Eq.(\ref{mixingatsneutralinos}), while $M_{\tilde{\nu_{j}}}$ is 
the mass of left-handed sneutrnos while $m_{\tilde{\chi}^{0}_{l}}$ is 
the mass of exchanged neutralinos. In the last line, $m_{\tilde{\chi}^{0}_{l}}$ is the 
neutralinos exchanged. The parameter $\theta_{\tilde{s}}$ is 
the mixing in sleptons defined at Eq.(\ref{mixingchargedleptons}), while 
$m_{s}$ is the mass of the exchanged lepton and $m_{\tilde{s}}$ 
is the mass of their respective slepton.

\section{Conclusions}
\label{sec:conclusion}

We presented the Supersymmetric version of the model presented 
at \cite{Machado:2017ksd} in the superfield formalism and we 
present also briefly some phenomenological consequences of this 
model. We have shown we have similar scalar potential as in the non-SUSY model, 
however in this Supersymmetric model there 
are no the couplings 
$\mu_{12}, \lambda_{9,10}$ then we can 
not have scotogenic mechanism, our 
neutrinos are Majorana particles. We 
also studied the mechanism to generate 
Majorana mass 
term at tree level to the right-handed neutrinos. The left-handed 
neutrinos get their masses at 1-loop 
level in this model, we show we still 
have the contribution arise from Figure 
1 from the non-SUSY model~
\cite{Machado:2017ksd}, but now we get 
two more contribution. The first 
one come from the neutralino-neutrino-sneutrino vertices drawn at Figure \ref{fig1}. The 
second arise from the charged 
lepton-neutrino-charged slepton 
interaction presented at Figure 
\ref{fig2}. We have shown a realistic radiative mechanism to generate masses to left-handed neutrinos with candidates for Dark Matter. In this model, the 
right-handed neutrinos and right-handed sneutrinos can be the Dark 
Matter candidate, it means this model has several 
possibilities for Dark Matter and we think it will be 
nice to study in more detail this subject.

\begin{center}
{\bf Acknowledgments} 
\end{center}
We would like to thanks to Instituto 
de F\'\i sica Te\'orica (IFT-Unesp) for their nice 
hospitality during the period I developed this article.

\appendix

\subsection{Scalar Potential}
\label{app:scalarpotential}

The first term at Eq.(\ref{vfdsoft}) is
\begin{equation}
V_{F}=\sum_{l}F_{l}^{\dagger }F_{l},
\end{equation}
where $l=H_{1,2}, D_{1,2}, D^{\prime}_{1,2}, \varphi , \phi$;
the $F$ terms are 
\begin{eqnarray}
F^{\dagger}_{H_{1}}&=&-\frac{\mu_{H}}{2}H_{2}, \,\
F^{\dagger}_{H_{2}}=-\frac{\mu_{H}}{2}H_{1}-
\frac{G^{l}_{ij}}{3}\tilde{L}_{iL}\tilde{E}_{jR},  \nonumber \\
F^{\dagger}_{D_{1}}&=&-\frac{\mu_{D_{1}}}{2}D^{\prime}_{1}, \,\
F^{\dagger}_{D^{\prime}_{1}}=-\frac{\mu_{D_{1}}}{2}D_{1}- 
\frac{G^{\nu}_{i}}{3}\tilde{L}_{iL}\tilde{N}_{1R}, \nonumber \\
F^{\dagger}_{D_{2}}&=&-\frac{\mu_{D_{2}}}{2}D^{\prime}_{2}, \,\
F^{\dagger}_{D^{\prime}_{2}}=-\frac{\mu_{D_{2}}}{2}D_{2}- 
\frac{F^{\nu}_{i \beta}}{3}\tilde{L}_{iL}\tilde{N}_{\beta R},  \nonumber \\
F^{\dagger}_{\varphi}&=&- \frac{H_{11}}{3}
\tilde{N}_{1R}\tilde{N}_{1R}, \,\
F^{\dagger}_{\phi}=- \frac{H_{\alpha \beta}}{3}
\tilde{N}_{\alpha R}\tilde{N}_{\beta R}.
\label{fterms}
\end{eqnarray}

The soft term that contribute to the scalar potential, see 
Eq.(\ref{lsoftint}), is given by
\begin{eqnarray}  
V_{soft}&=&- {\cal L}_{SMT}-{\cal L}_{Int}.
\label{soft}
\end{eqnarray}

The third term at Eq.(\ref{vfdsoft}) is
\begin{equation}
V_{D}=\frac{1}{2}\left[
D^{i}D^{i}+\left( D^{\prime}\right)^{2}\right],
\end{equation} 
where $i=1,2,3$. There is one $D$-term came from ${\cal L}_{Scalar}$ and from superpotential for each of the four gauge groups
\begin{eqnarray}
SU(2)_{L}: \,\ D^{i}&=&-\frac{g}{2} \left[
H_{1}^{\dagger}\sigma^{i}H_{1}+
H_{2}^{\dagger}\sigma^{i}H_{2}+
D^{\dagger}_{1}\sigma^{i}D_{1}+ D^{\prime \dagger}_{1}\sigma^{i}D^{\prime}_{1}+
D^{\dagger}_{2}\sigma^{i}D_{2}+ D^{\prime \dagger}_{2}\sigma^{i}D^{\prime}_{2}
\right], \nonumber \\
U(1)_{Y}: \,\ 
D^{\prime}=&-&\frac{g^{\prime}}{2}\left[
|H_{1}|^{2}-|H_{2}|^{2}+|D_{1}|^{2}-|D^{\prime}_{1}|^{2}+|D_{2}|^{2}-
|D^{\prime}_{2}|^{2}\right] ,
\label{dterms}
\end{eqnarray}
where $|H_{1}|^{2} \equiv H_{1}^{\dagger}H_{1}$ as usual.

%%%%%%%%%%%%%%%%%%%%%%%%%%%%%%%%%%%%%%%%%%%%%%%%%%%%%%%%%%%%%%%

\end{document}